%% file: main.tex
\documentclass[sigconf, 9pt, nonacm]{acmart}

\renewcommand\footnotetextcopyrightpermission[1]{}

\usepackage{makecell}
\usepackage{booktabs,tabularx,array}
\newcolumntype{Y}{>{\raggedright\arraybackslash}X}

\newcolumntype{N}[1]{>{\raggedright\arraybackslash}p{#1}}

\usepackage{amsmath,amssymb,amsfonts}
\usepackage{algorithmic}
\usepackage{graphicx}
\usepackage{textcomp}
\usepackage{xcolor}
\usepackage{adjustbox}
\usepackage{multirow}
\usepackage{pifont}
\usepackage{makecell} 
\usepackage[most]{tcolorbox}
\usepackage{enumitem}
\usepackage{xspace}

\newcommand{\Xmark}{\textcolor{red}{\ding{55}}}

\usepackage{booktabs}
\usepackage{siunitx}
\usepackage[table]{xcolor} 
\usepackage{tikz}          
\usepackage{tcolorbox}
\tcbuselibrary{listings}
\newtcolorbox{promptbox}[2][]{%
  colback=blue!5,        
  colframe=blue!60,      
  colbacktitle=blue!20,  
  coltitle=blue!50!black,        
  fonttitle=\bfseries,   
  fontupper=\ttfamily\small,   
  boxrule=0.6pt,         
  arc=2mm,               
  top=1mm, bottom=1mm,
  left=1mm, right=1mm,
  title=#2,#1
}

\newcommand{\ours}{$\mathsf{LockForge}$\xspace}

\begin{document}

\title{\textit{\ours}: Automating Paper-to-Code for Logic Locking with Multi-Agent Reasoning LLMs 
}
\vspace{-3mm}
\author{Akashdeep Saha$^\dagger$, Zeng Wang$^\ddagger$, Prithwish Basu Roy$^\ddagger$, Johann Knechtel$^\dagger$, \\ Ozgur Sinanoglu$^\dagger$, Ramesh Karri$^\ddagger$}
\affiliation{%
  \institution{$^\dagger$New York University Abu Dhabi}
  \city{Abu Dhabi}
  \country{UAE}
}

\affiliation{%
  \institution{$^\ddagger$New York University Tandon School of Engineering}
  \city{New York}
  \country{USA}
}
\email{{as19360,zw3464,pb2718,johann,ozgursin,rkarri}@nyu.edu}
\renewcommand{\shortauthors}{}
\begin{abstract}
Despite rapid progress in logic locking (LL), reproducibility remains a challenge as codes are rarely made public.
We present \ours, a first-of-its-kind, multi-agent large language model (LLM) framework that turns
LL
descriptions in papers into executable and tested code.
\ours provides a carefully crafted pipeline realizing forethought, implementation, iterative refinement, and a multi-stage validation, all to systematically bridge the gap
between prose and practice for complex LL schemes. For validation, we devise (i) an LLM-as-Judge stage with a scoring system
considering behavioral checks, conceptual mechanisms, structural elements, and reproducibility on benchmarks, and (ii) an independent LLM-as-Examiner stage for ground-truth assessment.
We apply \ours to 10 seminal LL schemes, many of which lack reference implementations.
Our evaluation on multiple SOTA LLMs, including ablation studies, reveals the significant complexity of the task.
We show that an advanced reasoning model and a sophisticated, multi-stage framework like \ours are required.
We release all implementations and benchmarks, providing a reproducible and fair foundation for evaluation of further LL research. 

\end{abstract}

\maketitle

\input{text/SectionI-Introduction}

\input{text/SectionII-Background}

\input{text/SectionIII-framework}

\input{text/SectionIV-similarity_score}
\input{text/SectionV-Experiments}

\input{text/SectionVI-Conclusion}
\bibliographystyle{ACM-Reference-Format}
\bibliography{bibliography}

\end{document}
\endinput

%% file: text/SectionI-Introduction.tex
\vspace{-3mm}
\section{Introduction}
In a globalized IC supply chain, cost and time-to-market pressures come with risks. These include IP piracy, counterfeiting, malicious  modifications, and reverse engineering. Logic locking (LL) addresses
these risks by inserting key-dependent logic, enabling correct IC behavior only for the right key(s).

\begin{figure}[!t]
    \centering
    \includegraphics[width=0.91\columnwidth, trim=0cm 0.5cm 0cm 0.2cm, clip]{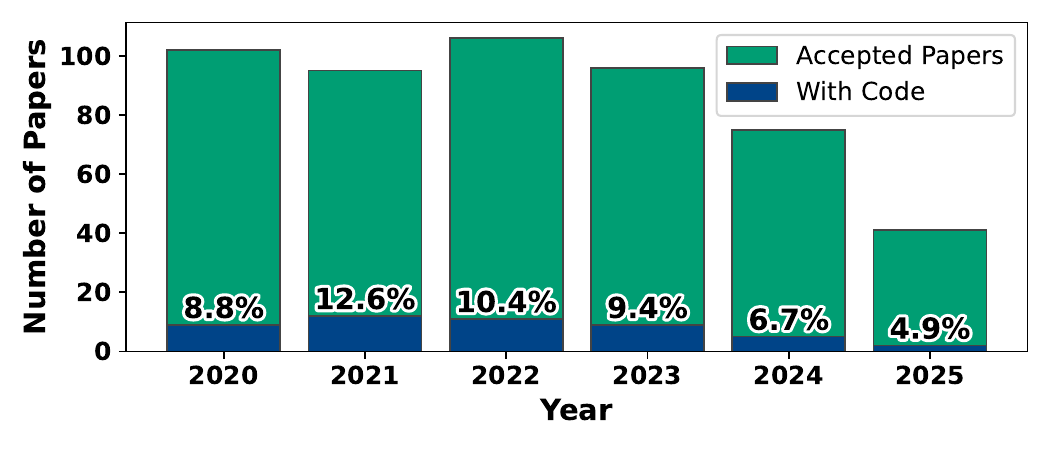} 
    \vspace{-2mm}
    \caption{Open source code for logic locking (2020--2025).}
    \label{fig:motivation}
    \vspace{-7mm}
\end{figure}

A major concern for (LL) research is the lack of reproducibility and independent evaluation. Figure~\ref{fig:motivation} shows that public codes are sparse, mostly $<$10\% of SOTA in the last five
years.\footnote{
	Obtained from https://openalex.org/ and
	https://github.com/} Likewise, in most research domains, reference implementations are rare, algorithmic details are distributed across papers, key experimental settings are missing, and data/tool
	access is restricted or proprietary~\cite{seo2025paper2code}. As a result, researchers have to painstakingly reconstruct methods from papers, slowing the community's progress and complicating independent
validation. Concurrently, large language models (LLMs) have advanced in natural language processing, including coding. Agentic LLMs can extract structure and concepts from technical PDFs, generate non-trivial codes, and repair it through execution-guided debugging and verification~\cite{openai2024gpt4,team2024gemini}.
 
In a timely and first-of-its-kind effort, we present \ours, a multi-agent LLM framework that converts LL papers into executable codes.
The \ours pipeline has four stages/phases as follows. Forethought ingests papers and extracts concepts.
Implementation performs prompted code synthesis with runtime feedback by compilation and local execution, also producing small-scale locked circuits. Refinement conducts concept-to-code alignment
via content mining, verifying that key mechanisms from the paper are realized in code, and runs smoke tests on the benchmark circuits. Validation employs independent agents to (i) probe conceptual fidelity and functional
correctness, checking the code explanations against the paper overview, through a rigorous scoring system, and (ii) pose reference-grounded true/false queries for final assessment.

We apply \ours to 10 diverse LL schemes~\cite{Shakya2020CASLock,Alaql2021SARO,Hashemi2025SRLL,Zhang2022TriLock,Maynard2023DKLock,Hu2024DECOR,Darjani2022ENTANGLE,Chakraborty2009HARPOON,Yasin2017SFLL,Wang2023AutoLock},
spanning point-function schemes, SAT-hard schemes, FSM-gated schemes, and membership-logic schemes, most lacking public implementations.
For end-to-end execution of generated codes on benchmarks, we observe consistent pass rates for simulation, equivalence checks, and our formal score criteria (which
verify all claimed paper features).
Ablations show that paper-grounded content mining/alignment, local execution, and independent examination all significantly contribute to fidelity and successful code execution.
We release all LL implementations, locked benchmarks, and checklists to support reproducibility and standardized comparative evaluation.
{Our contributions can be summarized as follows:}

\begin{itemize} [noitemsep]
    \item We propose \ours, a multi-agent, multi-stage LLM workflow with role isolation for LL coding and evaluation.
   Key to \ours is a formalized, objective, and reference-free validation of codes against papers through a systematic scoring system,
   coupled with agentic explanation alignment and paper-grounded true/false examination.

   \item We evaluate \ours on {10} LL schemes, most lacking public implementations, producing executable and verified codes.
   We open-source all LL implementations, locked benchmarks, and checklists to enable standardized and reproducible evaluation for the community at~\cite{lockforge_anonymous_repo}.

   \item  We benchmark multiple SOTA LLM backbones and run ablations on key stages.
	A key insight is that tackling this paper-to-code challenge for complex LL schemes requires not only a sophisticated agentic framework but also a SOTA reasoning engine, all carefully orchestrated.
    
\end{itemize}

%% file: text/SectionII-Background.tex
\section{Background}

\subsection{Logic Locking}
 
Early methods relied on random insertion of key gates for high corruption~\cite{roy2008epic,dacjv2012,tcadjv2013}, offering limited security guarantees.
The SAT attack~\cite{subramanyan2015evaluating} changed the goal of LL schemes from ad-hoc corruption to principled defenses, through SAT-hard structures~\cite{saha2020lopher}, controllable point
functions~\cite{yasin2016sarlock,Shakya2020CASLock,Yasin2017SFLL}, etc. A post-SAT scheme SFLL-HD~\cite{Yasin2017SFLL} introduced perturb-and-restore units.
TriLock~\cite{Zhang2022TriLock} allows tuning corruptibility while preserving resilience, DECOR~\cite{Hu2024DECOR} formalizes membership logic to permit multiple correct keys,
SARO~\cite{Alaql2021SARO} and AutoLock~\cite{Wang2023AutoLock} target scalability and automation, Entangle~\cite{Darjani2022ENTANGLE} couples perturb and restore networks.
DK-Lock~\cite{Maynard2023DKLock} and HARPOON~\cite{Chakraborty2009HARPOON} add counters or FSMs to decouple sequential activation from combinational behaviour. {\it Missing reference implementations for most schemes hinders independent evaluation.}

\begin{promptbox}[title={\textbf{Representative prompts used in \ours}}, halign title=center, fontupper=\footnotesize, fonttitle=\footnotesize]
 \scriptsize
\textbf{Forethoughts — LLM-A (PDF only)}
\begin{itemize}
  \item Read the paper and explain the logic locking scheme in it briefly.
  \item Can you list the crucial concepts of the logic locking scheme in this paper?
\end{itemize}
\textbf{Implementation — LLM-A}
\begin{itemize}
  \item Can you make a Python implementation of the logic locking and validate it using this input circuit? You should follow the paper's implementation exactly.
\end{itemize}
\textbf{Refinement Loop — LLM-A} 
\begin{itemize}
  \item Generate the YAML file for the BCSRP checklist for this logic locking scheme.
  \item Have you implemented all the concepts in your Python code?
  \item Validate your implementation with the BCSRP checklists.
\end{itemize}

\textbf{Validation — LLM-B and LLM-C (Code only)}
\begin{itemize}
  \item Explain the logic locking scheme implemented in this code.
  \item There are two descriptions of the logic locking scheme. Are these two explanations similar? \textit{(also ask LLM-A)}
  \item Use these BCSRP checklists in the YAML file to evaluate this implementation and score it (1 - 10).
  \item Generate 10 conceptual questions regarding the implementation with only true/false answers. \textbf{(prompt LLM-A)}
  \item Answer these questions with only T/F. 
\end{itemize}

\end{promptbox}

\subsection{LLM-Assisted Coding}

Early works on LLM-based coding targeted short snippets for isolated, toy programming tasks~\cite{austin2021program}.
As long-context reasoning and code synthesis have improved, the scope has shifted toward more demanding, end-to-end problems, using multi-agent or role-based workflows that mirror real-world development pipelines~\cite{jain2024livecodebench}.
{General-purpose AI platforms like~\cite{atlas_research} support data analysis, PDF processing, and interactive notebooks.}
In hardware security, LLMs are being recently used~\cite{wang2024llms,wang2025verileaky,saha2025gllamor}. {\it Ours is the first work to translate security schemes described in papers into codes for reproducibility.}

%% file: text/SectionIII-framework.tex
\begin{figure*}[th]
\centerline{\includegraphics[scale=0.34, trim={0cm 0.325cm 0.0cm 0.1cm},clip]{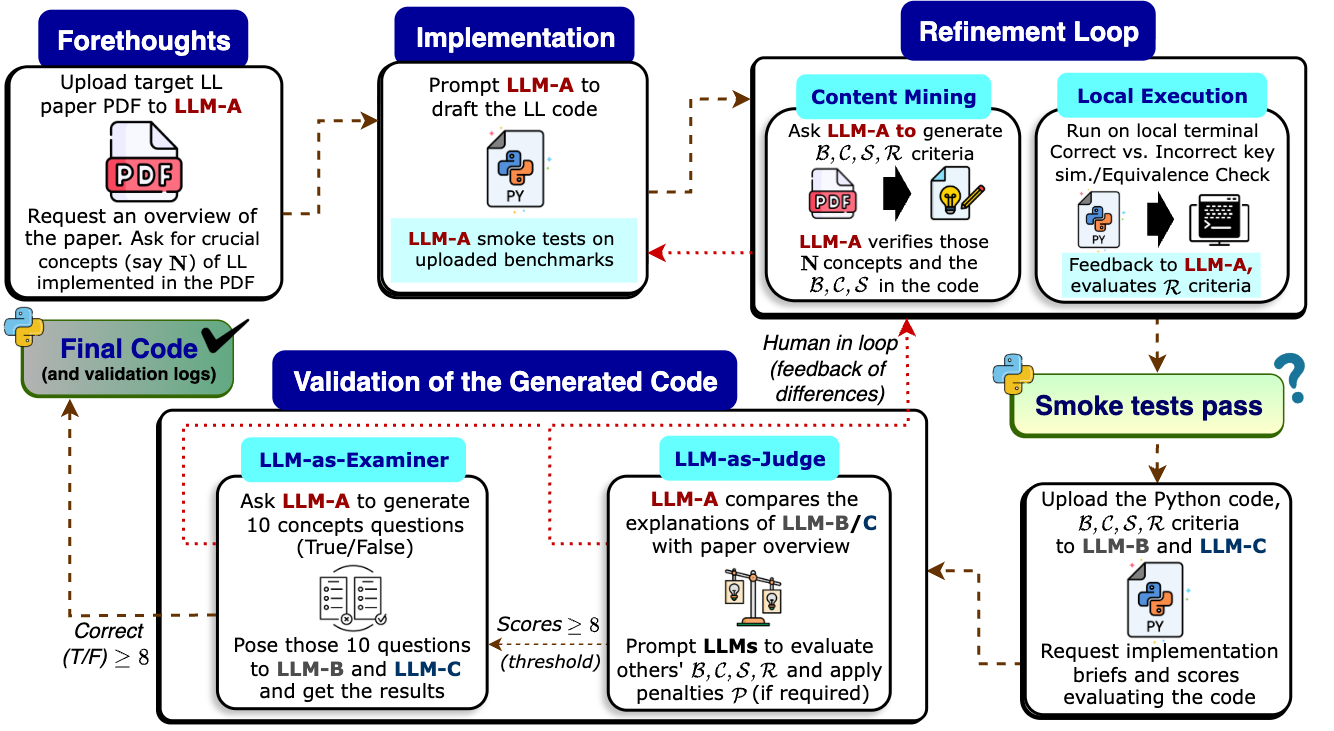}}
\caption{\ours is a multi-agent LLM workflow. It (i) parses a paper, (ii) drafts and refines code via concept mining and local execution/testing, and (iii) validates via independent LLMs and
	optional human assessment. {LLM-A is a coder with PDF access; LLM-B/C are judges/examiners with no PDF access.}}
\label{fig:framework} 
\vspace*{-0.1in}
\end{figure*}

\section{The \ours Framework}

\subsection{Motivation}

While a na\"ive, single-pass approach---uploading a paper and asking an LLM to `\textit{write the code}', may succeed for simple schemes~\cite{roy2008epic}, we find in exploratory experiments that this quickly fails
for contemporary LL works, due to the following challenges.
First, there is a significant semantic gap between narrative exposition and implementation, often because algorithmic details are spread throughout the paper, implementation parameters/settings are missing, test data is
unavailable,
etc.
Second, the context of LLMs is still limited when facing complex domains like LL~\cite{seo2025paper2code}.
While an LLM may readily compile some locked circuit, even with the correct key, it may not comprehend and implement the actual LL mechanisms.

To tackle these challenges,
we devise \ours,
a self-contained, multi-agent, multi-stage framework which implements LL schemes of various complexity that have no public reference codes.
\ours comprises supervised stages/tasks, for validation, and assigns them to specialized LLM roles. This staged workflow improves grounding, allows feedback,
and makes acceptance criteria explicit.

\subsection{Multi-Stage Pipeline}

Next, we describe the stages of \ours; refer Fig.~\ref{fig:framework}. Also see the list above for representative prompts.
 
\textbf{Forethoughts.}
This stage transforms unstructured text as obtained from the paper into implementation-level abstractions.
LLM-A ingests the PDF, produces a concise overview, enumerates core mechanisms, and derives artifact-level expectations. Thus, forethoughts
distills a persuasive narrative into a structured plan by extracting, say, $N$ concepts of the paper that are precise enough to guide coding and checks.
 
\textbf{Implementation.}
Next, LLM-A generates initial code, targeting benchmark formats and standard scripts.
Simulation on small circuits are run to ensure executability and reveal immediate errors, yielding self-contained codes ready for refinements.
  
\textbf{Refinement Loop.}
This stage first performs \textit{content mining}, by instructing the same LLM-A to generate the so-called $\mathcal{B,C,S,R,P}$ checklist for similarity scoring of the code against the paper.\footnote{%
$\mathcal{B}$ refers to behavioural checks, $\mathcal{C}$ to conceptual mechanisms, $\mathcal{S} $ to structural elements, and $\mathcal{R}$ to reproducibility on benchmarks.
Further, $\mathcal{P} $ is a non-compensatory penalty for critical omissions. 
See Sec.~\ref{sec:scoring} for more details.}
To do so, we initialize LLM-A with (i)~a definition of the checklist, (ii)~the scoring rules and computation, and (iii)~some templates for the actual checks. We instruct LLM-A to compile its findings into a simple YAML format.
Based on the latter, LLM-A is also instructed to iteratively patch the code until the $N$ derived concepts are realized and, thus, {the self-evaluation targets} for $\mathcal{B,C,S}$ are met (or until a feedback/iteration limit is reached).
Second, this stage performs \textit{local execution} on benchmarks to validate $\mathcal{R}$, including both simulation runs and equivalence checks.
As before,
feedback from this stage is returned to the same LLM-A for iterative revisions.
Finally, if all $\mathcal{B,C,S,R}$ thresholds are met, the code passes the smoke test and advances to the validation phase.

\textbf{Validation of the Generated Code.}
\ours certifies that generated codes implement the key mechanism described in LL papers even when no reference implementations exist.
We utilize role isolation---LLM-A reads the paper and performs iterative coding, while LLM-B and LLM-C analyze the code---along with similarity scoring and independent true/false
examination, as follows.

First, LLM-B and LLM-C independently judge, by summarizing and scoring the code on the $\mathcal{B,C,S,R,P}$ criteria, using the same YAML file compiled by LLM-A.
The aggregated and weighted criteria yield the final $\mathrm{Score}$ (Sec.~\ref{sec:scoring}, Eq.~\ref{eq:score}).
Only codes above an acceptance threshold, i.e., $\mathrm{Score}\geq 8$, advance.
Second, to avoid false positives by overfitting to the derived criteria, an independent examination of paper-grounded true/false statements is conducted.
For that, LLM-A derives a fixed set of statements from the paper, which LLM-B and LLM-C answer separately while inspecting only the code.
Codes advance for \#(correct true/false) $\geq 8$.
Finally, optional human-in-the-loop steps can audit the $\mathcal{B,C,S,R,P}$ checklists against the paper to complement automated scoring and review anomalies, if any, like low judge/examiner pass rates.

\textbf{Final Code.}
Passing validation proves that the paper's concepts are correctly realized and that the generated code executes as specified.
For reproducibility and independent verification, we release all LL implementations along with the $\mathcal{B,C,S,R,P}$ checklists~\cite{lockforge_anonymous_repo}.

\textbf{Summary.}
\ours engages multiple LLMs for different roles/tasks, which are operated independently but orchestrated via agentic interactions.
Content mining enables targeted code revisions, local execution runs help to cross-check the revisions and also uncover practical coding shortcomings like I/O issues, etc.
For validation, the LLM-as-Judge role captures 
breadth (concepts, structure, behaviour), while the LLM-as-Examiner role cross-checks for any missing semantics from the paper.

%% file: text/SectionIV-similarity_score.tex
\begin{table*}[t]
\centering
\caption{$\mathcal{B,C,S,P}$ checklists with brief descriptions for exemplary LL schemes}
\vspace{-5pt}
\label{tab:detailed_BCSRP_two}
\scriptsize
\renewcommand{\arraystretch}{.99}
\begin{adjustbox}{width=1.01\textwidth,center}
\begin{tabular}{|>{\centering\arraybackslash}m{0.85cm}|m{3.5cm}|m{5.6cm}|m{5.2cm}|m{3.5cm}|}

\hline
\textbf{Scheme} & \textbf{$\mathcal{B}$ (Behaviour)} & \textbf{$\mathcal{C}$ (Conceptual)} & \textbf{$\mathcal{S}$ (Structural)} & \textbf{$\mathcal{P}$ (Penalty; severity in brackets)} \\
\hline\hline
\textbf{CAS-Lock\cite{Shakya2020CASLock}} &
\textbf{Equivalence after insertion:} with the correct key, locked = golden (all inputs). 
\newline \textbf{Wrong-key corruption:} wrong key should yield output mismatches (some inputs). &
\textbf{Primary inputs keyed by XOR/XNOR:} each tapped signal randomly XOR/XNORed with a key bit. 
\newline \textbf{AND/OR cascade from keyed inputs:} keyed literals folded into 2-input serial cascades. 
\newline \textbf{Cascade output drives internal nets:} trigger $Y$ gates/perturbs selected nets. 
\newline \textbf{Cascade depth/fan-in match paper:} linear 2-input stages; intent depth $N{-}1$. &
\textbf{Key-gate count equals key bits:} inserted gates match declared key size. 
\newline \textbf{Cascade depth within expected range:} realized depth of $N{-}1$, comparable across chains. 
\newline \textbf{Cascade output reaches outputs:} live path from $Y$ to at least one primary output. 
\newline \textbf{No bypass under wrong key:} no alternate path restoring golden when $Y{=}1$. &
\textbf{Cascade absent (3):} no cascades detected. 
\newline \textbf{Cascade not connected (2):} trigger does not influence outputs. 
\newline \textbf{Key-bit count mismatch (1):} declared key size $\neq$ number of key gates. \\
\hline\hline
%
\textbf{Auto-\newline Lock\newline\cite{Wang2023AutoLock}} &
\textbf{Equivalence after insertion:} with the correct key, locked = golden (all inputs). 
\newline \textbf{Wrong-key corruption:} any incorrect key causes non-trivial deviation. 
\newline \textbf{Search loop integrates with netlist:} iterations read/modify circuit to score/apply changes. &
\textbf{Genotype defined:} candidate encodes edges/bits (e.g., $f_i,f_j,g_i,g_j$ with key). 
\newline \textbf{Fitness mixes corruption/overhead/constraints:} robustness, cost, validity combined. 
\newline \textbf{Iterative selection/mutation/crossover:} GA evolves population each generation. 
\newline \textbf{Insertion operator places key gates:} deterministic routine rewires chosen edges. &
\textbf{Inserted gate count matches budget:} equals requested key length. 
\newline \textbf{Insertion points valid in graph:} nets exist, distinct, and keep graph acyclic/legal. 
\newline \textbf{Fitness wired into search:} scores drive ranking and selection. 
\newline \textbf{No bypass under wrong key:} only keyed path restores correct behaviour. &
\textbf{Insertion operator absent (3):} no effective rewiring. 
\newline \textbf{Fitness unused (2):} score computed but not applied (random search used). 
\newline \textbf{Budget violation (1):} too many/few key gates relative to target. \\
\hline
\end{tabular}
\end{adjustbox}
\vspace{-.3cm}
\end{table*}

\subsection{Similarity Scoring}
\label{sec:scoring}

The similarity $\mathrm{Score} \in [1,10]$ is defined in Eq.~\ref{eq:score}. It is computed from the $\mathcal{B,C,S,R,P}$ criteria using
multi-criteria decision analysis (MCDA)~\cite{hwang1981methods}.\footnote{
MCDA is a family of methods that converts performance on multiple criteria into a single ranking by normalizing, weighting, and aggregating those criteria linearly. We apply a small non-compensatory penalty so that
\textit{hard failures} cannot be averaged away, realizing a veto-style safeguard~\cite{roy1991outranking}.}
As indicated, we consider four criteria: behavioural checks $\mathcal{B}$, conceptual mechanisms $\mathcal{C}$, structural elements $\mathcal{S}$, and reproducibility $\mathcal{R}$. Details are provided next, and
Tab.~\ref{tab:detailed_BCSRP_two} provides the $\mathcal{B,C,S,R,P}$ checklists with rationales for exemplary LL schemes.\footnote{%
Checklists for all other LL schemes are provided in our release~\cite{lockforge_anonymous_repo}.}
 
\begin{align}
\label{eq:score}
\mathrm{Score}(x)
&= \max\Big\{\,1,\ 
\left\lfloor 10\Big(0.4\,\mathcal{B}(x)+0.3\,\mathcal{C}(x) +\,0.2\,\mathcal{S}(x) \right. \\[-2pt]
&\qquad\left. +0.1\,\mathcal{R}(x)\Big)\right\rfloor
- \mathcal{P}(x)\Big\} \notag
\end{align}
Using empirical insights and without loss of generality, we use weights of 0.4, 0.3, 0.2, and 0.1, respectively, for $\mathcal{B,C,S,R}$.

\textbf{Behavioural checks $\mathcal{B}$} quantifies the ratio of verification checks satisfied,
including correct-key equivalence to a golden model, wrong-key behaviour, and temporal/FSM semantics. 

\textbf{Conceptual mechanisms $\mathcal{C}$} measures the paper-specific mechanisms realized in code.
That is,
$N$ core concepts are extracted from the paper, defining set $\mathcal{F}_{\mathrm{req}}$, and their implementations are verified against the code ($\mathcal{F}_{\mathrm{impl}}$).
\begin{equation}
\mathcal{C}(x) = \frac{\big|\mathcal{F}_{\mathrm{impl}}(x)\cap \mathcal{F}_{\mathrm{req}}\big|}
             {\big|\mathcal{F}_{\mathrm{req}}\big|}.
\label{eq:C}
\end{equation}

\textbf{Structural elements $\mathcal{S}$} quantifies the share of essential structural elements present and correctly implemented, e.g., key gates, restore/perturb modules, comparators, and FSM states.
\begin{equation}
\mathcal{S}(x) = 1 - \frac{\big|\mathcal{M}(x)\big|}{\big|J\big|}
\label{eq:S}
\end{equation}
where $J$ is the set of structural components considered,
and $\mathcal{M}(x)\subseteq J$ covers the implementation mismatches in $x$.

\textbf{Reproducibility $\mathcal{R}$} measures fidelity of code execution on paper-specified benchmarks.
$\mathcal{R}$ rates end-to-end successful runs, runs with minor code tweaks, and fails as 1.0, 0.5, 0.0.

\textbf{Penalty $\mathcal{P}$} is the sum of severity grades for critical omissions, with each minor, major, and severe omission rated as 1, 2, and 3, respectively. Note that the categorization is based on the specifics
for the LL scheme as derived by LLM-A.

%% file: text/SectionV-Experiments.tex
\section{Experimental Investigation}

\subsection{Setup}

\textbf{LL Schemes and Benchmarks.}
We apply \ours to 10 LL schemes~\cite{Shakya2020CASLock,Alaql2021SARO,Hashemi2025SRLL,Zhang2022TriLock,Maynard2023DKLock,Hu2024DECOR,Darjani2022ENTANGLE,Chakraborty2009HARPOON,Yasin2017SFLL,Wang2023AutoLock},
covering different approaches and implementation styles for LL.
We use benchmarks from these papers;  ISCAS-85, ITC-99, and MCNC are common to many schemes. We utilize key-bit sizes as specified in these papers.

\textbf{Models and Roles.} We benchmark ChatGPT-5 Thinking, DeepSeek-V3, and Gemini-2.5pro in combinations of distinct LLM-A/B/C roles. An LLM may behave differently as coder vs evaluator and cross-model judging reduces single-model bias. Without loss of generality, we apply thresholds $\mathrm{Score}\!\ge\!8$ and T/F $\!\ge\!8$, and a budget of 10 iterations for refinement.

\textbf{Metrics.} We report $\mathcal{B,C,S,R,P}$, the derived $\mathrm{Score}$, T/F accuracy, and PASS/INCORRECT/FAIL for local execution.
PASS indicates that the generated code runs successfully and matches the expected outputs for correct vs wrong keys on at least 90\% of benchmarks.
INCORRECT indicates that the code runs but outputs deviate significantly for
one or more key settings/benchmarks. FAIL indicates code failing to execute.

\subsection{Results}

\begin{figure}[]
    \centering
    \includegraphics[width=1.1\columnwidth, trim=0.4cm 0.7cm 0cm 0.6cm, clip]{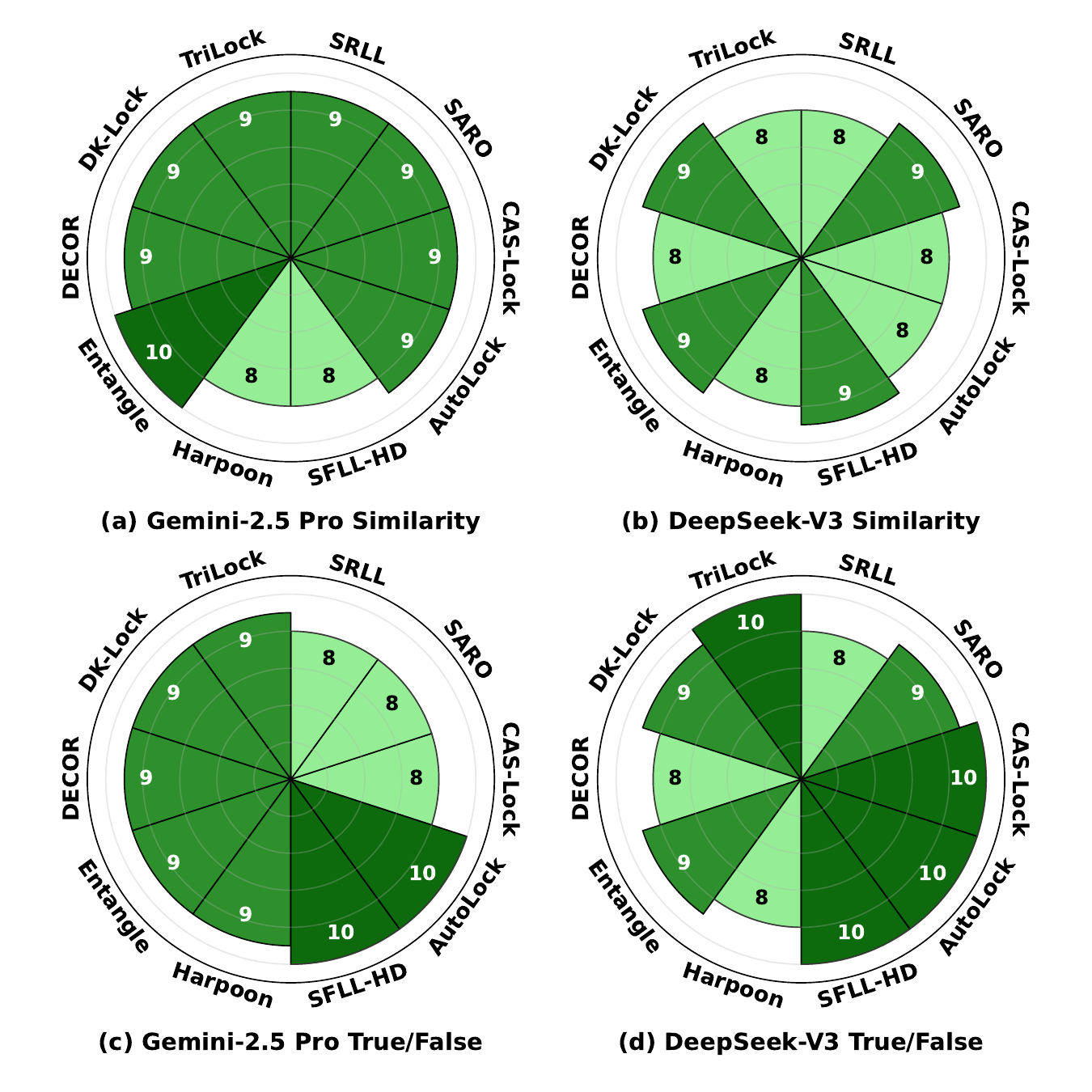} 
    \vspace{-5mm}
    \caption{Scoring ChatGPT-5 codes by Gemini-2.5pro and DeepSeek-V3.}
    \vspace{-5mm}
    \label{fig:ChatGPT_piechart}
\end{figure}

\textbf{Overall Performance.}
Shown in Fig.~\ref{fig:ChatGPT_piechart},
ChatGPT-5 acting as coder achieves the highest similarity scores, best examination pass rate, and
lower $\mathcal{P}$ (shown later), across all 10 LL schemes.
In contrast, DeepSeek-V3 and Gemini-2.5pro acting as coders require more refinement and saturate earlier (also shown later).
\emph{ \underline{Takeaways:} } Thorough reasoning, enabled by the latest ChatGPT-5 model integrated into the agentic pipeline of \ours, yields better code quality.

\begin{figure}[t]
    \centering
    \includegraphics[width=1.05\columnwidth, trim=0.26cm 0.32cm 0.3cm 0.25cm, clip]{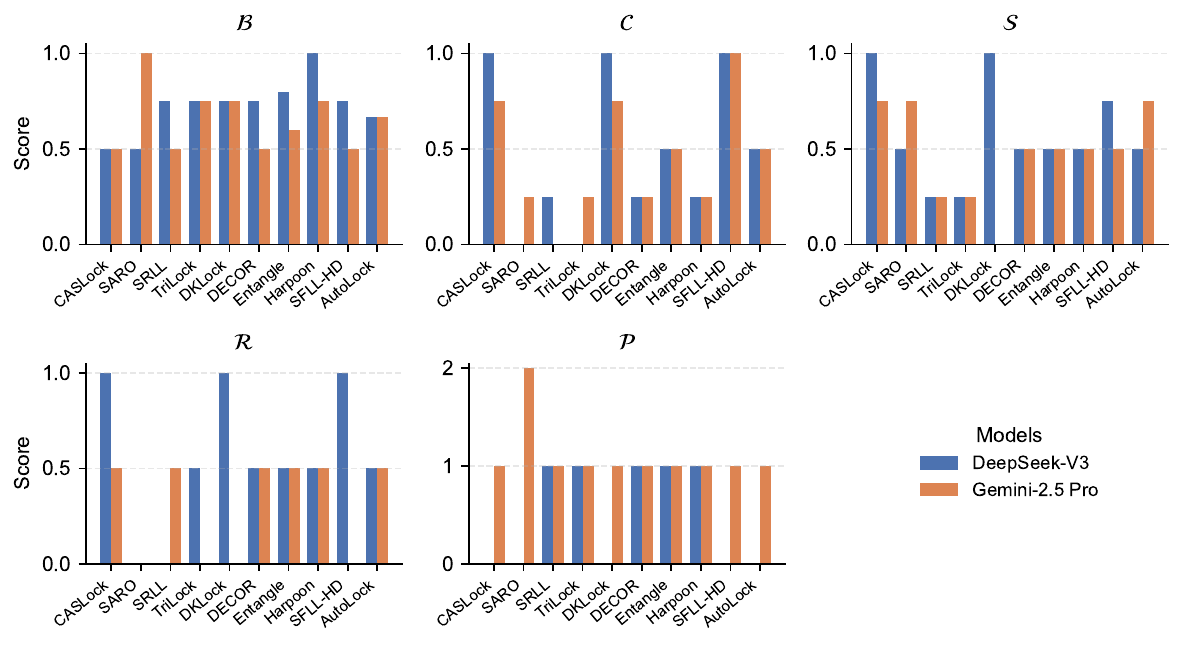} 
    \caption{Gemini-2.5pro, DeepSeek-V3 coding; 
    evaluation by ChatGPT-5.}
    \label{fig:ge-ds-bcsrp}
\end{figure}

\begin{figure}[t]
    \centering
    \includegraphics[width=1.0\columnwidth, trim=0cm 0.17cm 0cm 0.25cm, clip]{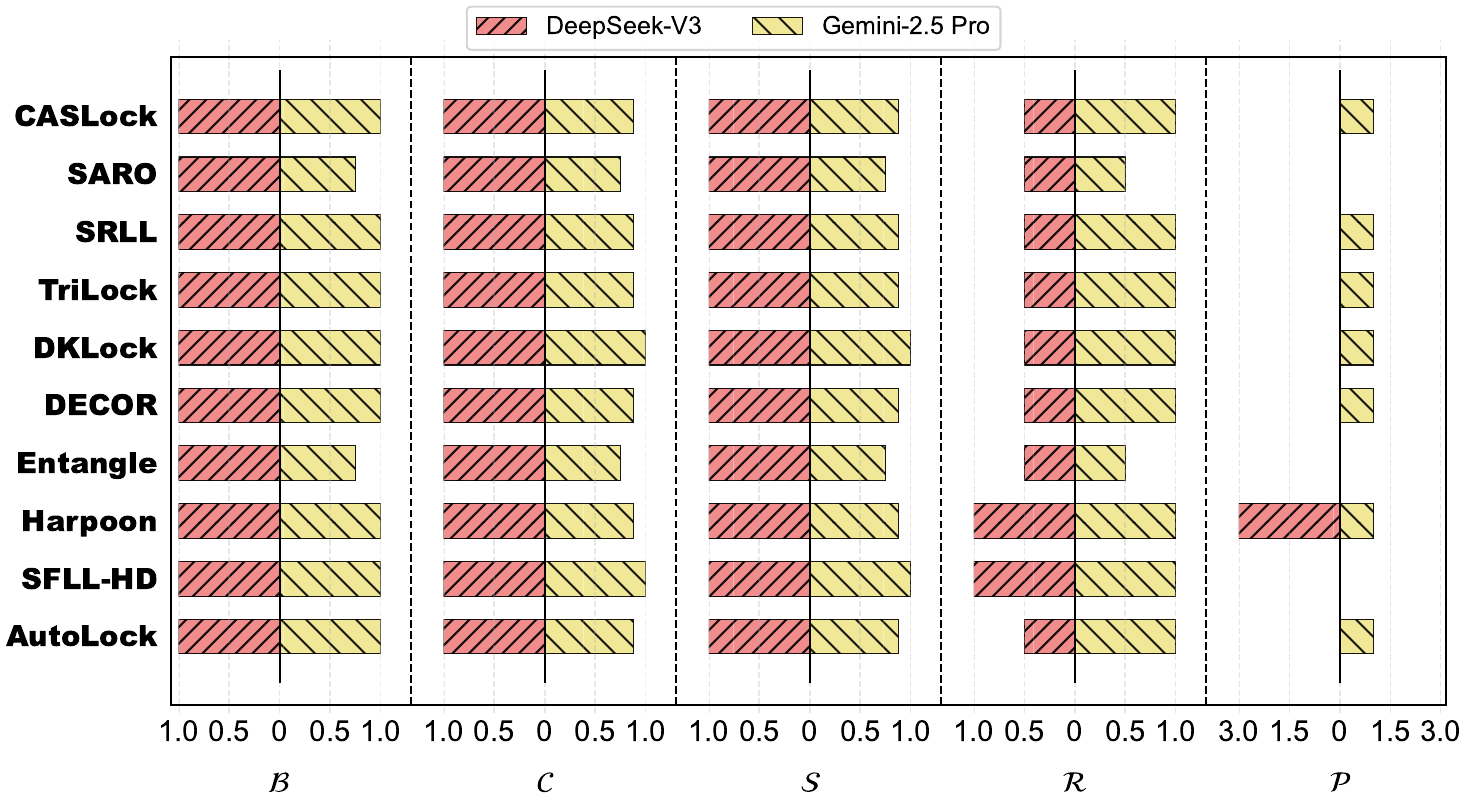} 
    \caption{ChatGPT-5 coding;
    evaluation by Gemini-2.5pro, DeepSeek-V3.}
    \label{fig:gpt-bcsrp}
    \vspace{-5mm}
\end{figure}

\textbf{Scoring Details for Cross-Model Evaluation.}
Figure~\ref{fig:ge-ds-bcsrp} shows the $\mathcal{B,C,S,R,P}$ scoring for DeepSeek-V3 and Gemini-2.5pro generated codes as evaluated by ChatGPT-5.
Vice versa, Fig.~\ref{fig:gpt-bcsrp} shows scoring for ChatGPT-5 generated codes as evaluated by DeepSeek-V3 and Gemini-2.5pro.
Again, ChatGPT-5 codes consistently show the highest scoring, with minimal $\mathcal{P}$, and are reproducible. \emph{ \underline{Takeaways:} }
The significant performance gap between coders highlights the task's complexity---success is not just about code generation, but about high-fidelity conceptual understanding, a trait only observed in the latest reasoning models. For fair evaluation, we do not use same models as coder, judge, and examiner; cross-model evaluation curbs any model-specific `blind spots'.

\textbf{Coding Refinement.}
Table~\ref{tab:Deepseek_table} details coding outcomes for DeepSeek-V3,
with ChatGPT-5 and Gemini-2.5 Pro each serving as judge and examiner.
Notably, only two implementations pass local execution.
Even after iterative agentic feedback, various crucial concepts of most LL schemes were not realized by DeepSeek-V3.
Likewise, Tab.~\ref{tab:Gemini_table} details coding outcomes for Gemini-2.5pro, where none of the implementations realize fully correct code before exhausting the 10-iteration limit.
Refinement often saturates early, with no further gains observed for $\mathcal{B,C,S,R}$ scorings.
In contrast, codes generated by ChatGPT-5 always pass local execution, with only minor implementation issues, if any.
\emph{\underline{Takeaways:} } Refinement cannot compensate for concept omissions. {Content mining} and quality of the checklist, which informs the concepts to be implemented, are a first-order priority.
Clearly,  thorough reasoning is crucial to fully understand the semantics of the LL domain.

\begin{figure}[tb]
    \centering
    \includegraphics[width=1.0\columnwidth, trim=0.4cm 0.8cm 0cm 1.99cm, clip]{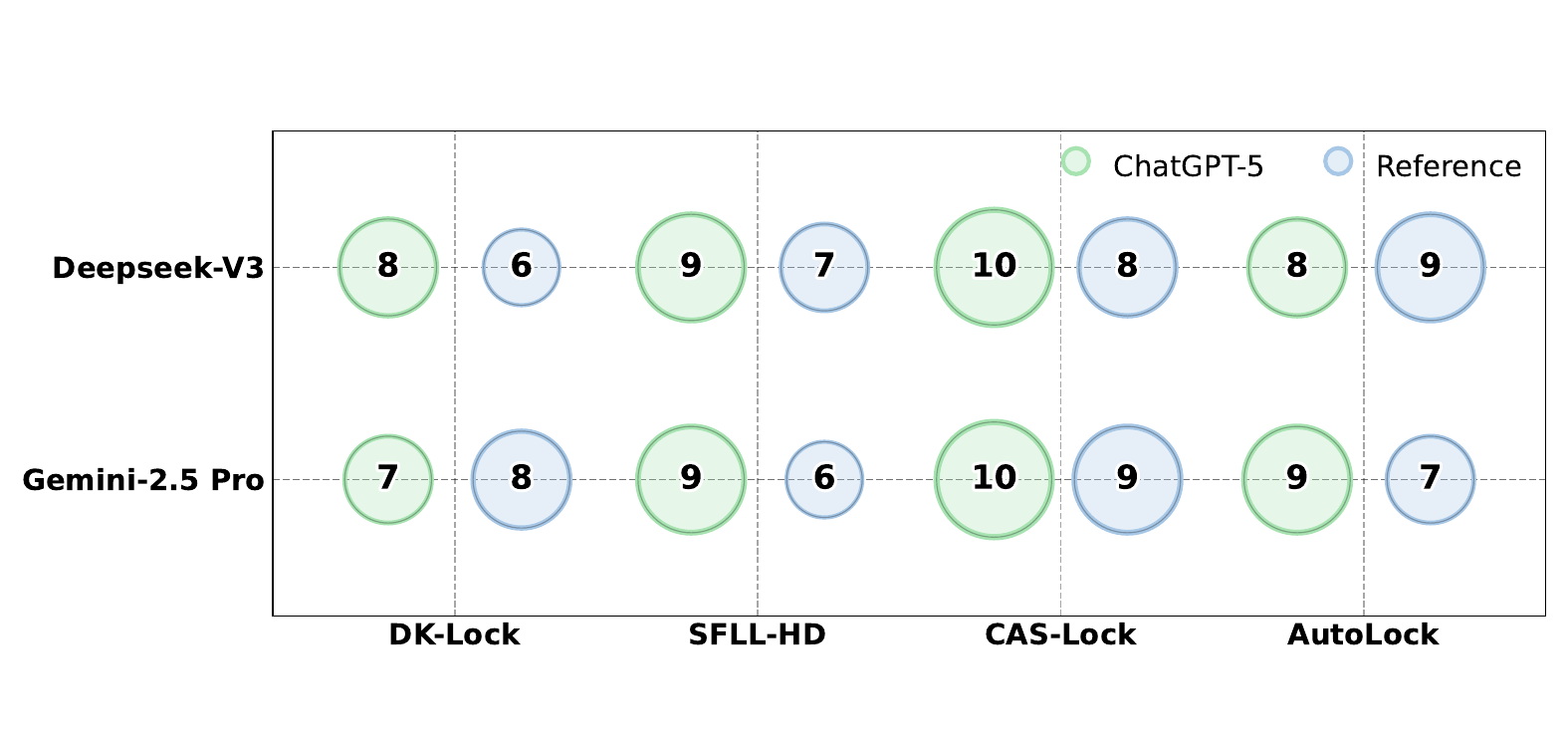} 
    \vspace{-0.7cm}
    \caption{Overall scores (computed via Eq.~\ref{eq:score}) comparing ChatGPT-5–generated code with available reference codes.}
    \label{fig:gpt_VS_local}
\end{figure}

\textbf{Comparison with Reference Implementations.}
Most LL schemes offer no publicly available implementations. For LL schemes with code released,\footnote{%
DK-Lock at \url{https://github.com/cars-lab-repo/DKL}, SFLL-HD at \url{https://github.com/circuitgraph/logiclocking},
and CAS-Lock and AutoLock via \cite{lockforge_anonymous_repo}.}
we compare them to \ours using ChatGPT-5 coder in Fig.~\ref{fig:gpt_VS_local}.
\emph{ \underline{Takeaways:} } \ours performs on par with reference codes, confirming the faithful nature of generated artifacts.

\textbf{Comparison with Related Work.}
ATLAS~\cite{atlas_research} and P2C~\cite{seo2025paper2code} are commercial and academic offerings, respectively, for paper-to-code tasks. We followed their setup instructions to implement four  LL schemes.
As Tab.~\ref{tab:platform_scores} shows, both underperformed. ATLAS often produced tangential code (misidentified mechanisms, pattern-matched scaffolds) and lacked hardware-awareness.
P2C was paper-focused yet struggled with hardware-centric aspects, resulting in incomplete netlist integration. Notably, P2C handled the ML-based scheme AutoLock~\cite{Wang2023AutoLock} relatively well, as expected due to its ML-focused pipeline~\cite{seo2025paper2code}.
\emph{\underline{Takeaways:}} The complex task of paper-to-code for LL requires hardware-aware reasoning and agentic feedback.
 
\begin{table}[tb]
\centering
\scriptsize
\caption{ChatGPT-5 evaluation scores for related works.}
\label{tab:platform_scores}
\begin{adjustbox}{width=0.9\columnwidth}
\begin{tabular}{|l|c|c|c|c|}
\hline
\textbf{Platform} & \textbf{SARO} & \textbf{AutoLock} & \textbf{Harpoon} & \textbf{DK-Lock} \\
\hline
\hline
\textbf{ATLAS~\cite{atlas_research} } & 2/10 & 2/10 & 4/10 & 1/10 \\
\hline
\textbf{P2C~\cite{seo2025paper2code} }   & 3/10 & 7/10 & 3/10 & 4/10 \\
\hline
\end{tabular}
\end{adjustbox}
\end{table}

\definecolor{r1}{HTML}{FDE2E2} 
\definecolor{r2}{HTML}{EF9A9A} 
\definecolor{r3}{HTML}{C62828} 

\definecolor{y4}{HTML}{FFFDE7} 
\definecolor{y5}{HTML}{FFF9C4} 
\definecolor{y6}{HTML}{FFF59D} 
\definecolor{y7}{HTML}{FFF176} 

\definecolor{sea8}{HTML}{D1F5F0}
\definecolor{sea9}{HTML}{76D7C4}
\definecolor{sea10}{HTML}{1F9E89}

\newcommand{\tfcell}[1]{%
  \begingroup
  \count0=#1\relax
  \ifnum\count0=1
    \cellcolor{r1}\bfseries #1%
  \else\ifnum\count0=2
    \cellcolor{r2}\bfseries #1%
  \else\ifnum\count0=3
    \cellcolor{r3}\color{white}\bfseries #1%
  \else\ifnum\count0=4
    \cellcolor{y4}\bfseries #1%
  \else\ifnum\count0=5
    \cellcolor{y5}\bfseries #1%
  \else\ifnum\count0=6
    \cellcolor{y6}\bfseries #1%
  \else\ifnum\count0=7
    \cellcolor{y7}\bfseries #1%
  \else\ifnum\count0=8
    \cellcolor{sea8}\bfseries #1%
  \else\ifnum\count0=9
    \cellcolor{sea9}\bfseries #1%
  \else
    \cellcolor{sea10}\color{white}\bfseries #1%
  \fi\fi\fi\fi\fi\fi\fi\fi\fi
  \endgroup
}

%
%
\setlength{\arrayrulewidth}{0.6pt}
\renewcommand{\arraystretch}{1.2}
\newcolumntype{P}[1]{>{\raggedright\arraybackslash}p{#1}}

\definecolor{r0}{HTML}{8E0000} 
\definecolor{r1}{HTML}{B71C1C} 
\definecolor{r2}{HTML}{EF9A9A} 
\definecolor{r3}{HTML}{FDE2E2} 

\definecolor{y4}{HTML}{FFFDE7}
\definecolor{y5}{HTML}{FFF9C4}
\definecolor{y6}{HTML}{FFF59D}
\definecolor{y7}{HTML}{FFF176}

\definecolor{sea8}{HTML}{D1F5F0}
\definecolor{sea9}{HTML}{76D7C4}
\definecolor{sea10}{HTML}{1F9E89}

\newcommand{\scorecell}[1]{%
  \begingroup
  \def\val{#1}%
  \ifdim \val pt < 1pt
    \cellcolor{r0}\color{white}\bfseries #1%
  \else\ifdim \val pt < 2pt
    \cellcolor{r1}\color{white}\bfseries #1%
  \else\ifdim \val pt < 3pt
    \cellcolor{r2}\bfseries #1%
  \else\ifdim \val pt < 4pt
    \cellcolor{r3}\bfseries #1%
  \else\ifdim \val pt < 5pt
    \cellcolor{y4}\bfseries #1%
  \else\ifdim \val pt < 6pt
    \cellcolor{y5}\bfseries #1%
  \else\ifdim \val pt < 7pt
    \cellcolor{y6}\bfseries #1%
  \else\ifdim \val pt < 8pt
    \cellcolor{y7}\bfseries #1%
  \else\ifdim \val pt < 9pt
    \cellcolor{sea8}\bfseries #1%
  \else\ifdim \val pt < 10pt
    \cellcolor{sea9}\bfseries #1%
  \else
    \cellcolor{sea10}\color{white}\bfseries #1%
  \fi\fi\fi\fi\fi\fi\fi\fi\fi
  \endgroup
}
\newcommand{\simcellpass}{\textcolor{green!50!black}{\ding{51}}~\textbf{PASS}} 
\newcommand{\simcellfail}{\textcolor{red!70}{\ding{55}}~\textbf{FAIL}}         
\newcommand{\simcellinc}{\textcolor{orange!80!black}{$\triangle$}~\textbf{INCORR}}

\begin{table*}[t]
\centering
\caption{Ablation results. Scoring by Gemini-2.5pro and DeepSeek-V3 on ChatGPT-5 generated codes. }
\label{tab:CM_LE_ablation}
\scriptsize
\renewcommand{\arraystretch}{1.2}
\begin{adjustbox}{width=1.0\textwidth}
\begin{tabular}{|l|c|c|l|c|c|l|}
\hline
\multirow{2}{*}{\textbf{LL Scheme}} &
\multicolumn{3}{c|}{\textbf{Content Mining Ablation}} &
\multicolumn{3}{c|}{\textbf{Local Execution Ablation}} \\
\cline{2-7}
 & \textbf{Gemini} & \textbf{DeepSeek} & \textbf{Common Remarks} & \textbf{Gemini} & \textbf{DeepSeek} & \textbf{Common Remarks} \\
\hline
\textbf{CAS-Lock~\cite{Shakya2020CASLock}} & \scorecell{2} & \scorecell{2} &
\begin{tabular}[t]{@{}l@{}}missing dual cascades, complementary patterns\\
single cascade only; fixed AND-only cascade\end{tabular} &
 \scorecell{7} & \scorecell{7} &
\begin{tabular}[t]{@{}l@{}}Output misalignment (\texttt{\_LOCKED\_wire}/name) reuse; bench-specific;\\
fixed key for all\end{tabular} \\
\hline
\textbf{DECOR~\cite{Hu2024DECOR}} & \scorecell{2} & \scorecell{3} &
\begin{tabular}[t]{@{}l@{}}Single correct key; DECOR structural transform; \texttt{is\_correct} logic;\\
output correction mechanism\end{tabular} &
 \scorecell{5} & \scorecell{6} &
\begin{tabular}[t]{@{}l@{}}Local execution step absent; conceptual-only check; reproducibility unstable;\\
several benches show mismatches\end{tabular} \\
\hline
\textbf{DK-Lock~\cite{Maynard2023DKLock}} & \scorecell{4} & \scorecell{3} &
\begin{tabular}[t]{@{}l@{}}M-cycle activation requirement; sequential counter; simple XOR locking;\\
output clamping; M-cycle activation counter\end{tabular} &
 \scorecell{6} & \scorecell{7} &
\begin{tabular}[t]{@{}l@{}}Keys auto-padded/truncated, hence, mismatches despite ``correct'' JSON;\\
output reordering can flip results\end{tabular} \\
\hline
\textbf{TriLock~\cite{Zhang2022TriLock}} & \scorecell{3} & \scorecell{4} &
\begin{tabular}[t]{@{}l@{}}Structural obfuscation; EF trigger; no ES/EF mechanisms;\\
no state obfuscation; no $k_s/k_f$ separation\end{tabular} &
 \scorecell{5} & \scorecell{6} &
\begin{tabular}[t]{@{}l@{}}Key pin order not exported, hence, ``correct key'' mismatches\end{tabular} \\
\hline
\textbf{SRLL~\cite{Hashemi2025SRLL}} & \scorecell{5} & \scorecell{6} &
\begin{tabular}[t]{@{}l@{}}No CS calculation; no alternative blocks; no combinational-cycle insertion;\\
no LUT-based withholding with TT keys\end{tabular} &
 \scorecell{6} & \scorecell{6} &
\begin{tabular}[t]{@{}l@{}}Local run inconsistent per-bench (some fail); key format mismatch (JSON expected);\\
script prints CSV, parser fail\end{tabular} \\
\hline
\textbf{Entangle~\cite{Darjani2022ENTANGLE}} & \scorecell{4} & \scorecell{4} &
\begin{tabular}[t]{@{}l@{}}No signal entanglement; no dummy logic/cones; faulty restore unit;\\
external-signal entanglement\end{tabular} &
 \scorecell{5} & \scorecell{7} &
\begin{tabular}[t]{@{}l@{}}Sim accepted incorrect/correct keys across random inputs;\\
no golden/locked equivalence check done\end{tabular} \\
\hline
\textbf{SARO~\cite{Alaql2021SARO}} & \scorecell{2} & \scorecell{2} &
\begin{tabular}[t]{@{}l@{}}Rename-then-wrap; avoids cycling; no correct key identity guarantee;
XOR corruption\end{tabular} &
 \scorecell{3} & \scorecell{4} &
\begin{tabular}[t]{@{}l@{}}Not generic to all bench formats; local run skipped\end{tabular} \\
\hline
\textbf{Harpoon~\cite{Chakraborty2009HARPOON}} & \scorecell{4} & \scorecell{6} &
\begin{tabular}[t]{@{}l@{}}Missing sequential entrance lock; no clear obfuscate/authenticate phases;\\
no state-conditioned corrupt/restore; no reuse of unreachable states\end{tabular} &
 \scorecell{5} & \scorecell{7} &
\begin{tabular}[t]{@{}l@{}}Local execution failed due to parsing; reproducibility not evaluated across benches\end{tabular} \\
\hline
\textbf{AutoLock~\cite{Wang2023AutoLock}} & \scorecell{3} & \scorecell{2} &
\begin{tabular}[t]{@{}l@{}}Attack-in-loop fitness; cycle, conflict detection \& repair;\\
genotype with per-gene embedding\end{tabular} &
 \scorecell{5} & \scorecell{4} &
\begin{tabular}[t]{@{}l@{}}Correct key mismatch from key-order mismatch; no wrong-key sims or miter checks\end{tabular} \\
\hline
\textbf{SFLL-HD~\cite{Yasin2017SFLL}} & \scorecell{5} & \scorecell{6} &
\begin{tabular}[t]{@{}l@{}}No stripped functionality; incorrect restore unit\end{tabular} &
 \scorecell{7} & \scorecell{6} &
\begin{tabular}[t]{@{}l@{}}Correct key check sample-based only (no miter); occasional small mismatches possible\end{tabular} \\
\hline
\end{tabular}
\end{adjustbox}
\vspace{-.2cm}
\end{table*}
 
\begin{table}[]
\centering
\caption{DeepSeek-V3 codes; {Gemini-2.5pro}, {ChatGPT-5} evaluate. Scores are binned. Generated remarks cross-verified.}
\label{tab:Deepseek_table}
\begin{adjustbox}{width=0.97\columnwidth}
\begin{tabular}{|c|c|c|c|P{0.50\linewidth}|}
\hline 
\multicolumn{1}{|c|}{\textbf{Logic Locking}} &
\multicolumn{2}{c|}{\textbf{Similarity Check (1–10)}} &
\multicolumn{1}{c|}{\textbf{Local}} &
\multicolumn{1}{c|}{\textbf{Common remarks from }} \\
\cline{2-3}
\textbf{Schemes} & \textbf{Gemini-2.5} & \textbf{ChatGPT-5} & \textbf{Simulation?} & \textbf{\hspace{0.45cm} Gemini-2.5 and ChatGPT-5}\\
\hline \hline
\textbf{CAS-Lock}~\cite{Shakya2020CASLock} & \scorecell{8}   & \scorecell{8}   & \simcellpass & Initial cascade wrong; same lock gates for all inputs. Corrected with feedback. \\
\hline
\textbf{SARO}~\cite{Alaql2021SARO}     & \scorecell{2}   & \scorecell{3}   & \simcellfail & Concepts not implemented. No Rename-then-Wrap. Incorrect T3 transform implementation. \\
\hline
\textbf{SRLL}~\cite{Hashemi2025SRLL}     & \scorecell{3}   & \scorecell{3}   & \simcellfail & Wrong implementation; No LUT-based withholding w/ truth-table keys, key-controlled entanglement with cone-external signals. Random XOR/XNOR LL. \\
\hline
\textbf{TriLock}~\cite{Zhang2022TriLock}  & \scorecell{5}   & \scorecell{3}   & \simcellinc  & Concepts not implemented. No temporal ES/phase gating; no k-deep/EF modeling. \\
\hline
\textbf{DK-Lock}~\cite{Maynard2023DKLock}  & \scorecell{9}   & \scorecell{9}   & \simcellpass & Most  crucial concepts implemented. \\
\hline
\textbf{DECOR}~\cite{Hu2024DECOR}    & \scorecell{4}   & \scorecell{4} & \simcellinc  & inserts XOR/XNOR key-gates; does not implement UDC/cofactor altering; no is\_correct+G/L correction logic. \\
\hline
\textbf{Entangle}~\cite{Darjani2022ENTANGLE} & \scorecell{5}   & \scorecell{5}   & \simcellinc  & Incorrect control logic and restore unit implementation. \\
\hline
\textbf{Harpoon}~\cite{Chakraborty2009HARPOON}  & \scorecell{6} & \scorecell{5}   & \simcellinc  & No state-space expansion implemented. \\
\hline
\textbf{SFLL-HD}~\cite{Yasin2017SFLL}  & \scorecell{8} &   \scorecell{8}  & \simcellinc  & FSC built via truth-table enumeration. Simulation fails on large circuits.\\
\hline
\textbf{AutoLock}~\cite{Wang2023AutoLock} & \scorecell{6}   & \scorecell{6}   & \simcellinc  & wrong fitness function implementation. Locking insertion flawed. \\
\hline
\end{tabular}
\end{adjustbox}
\vspace{-.2cm}
\end{table}
%
%
%
\providecommand{\simcellna}{\cellcolor{black!10}{\textbf{N/A}}}
\begin{table}[]
\centering
\caption{Gemini-2.5pro: codes;  DeepSeek-V3, ChatGPT-5: evaluate. Scores are binned. Generated remarks cross-verified.}
\label{tab:Gemini_table}
\begin{adjustbox}{width=0.97\columnwidth}
\begin{tabular}{|c|c|c|c|P{0.50\linewidth}|}
\hline
\multicolumn{1}{|c|}{\textbf{Logic Locking}} &
\multicolumn{2}{c|}{\textbf{Similarity Check (1–10)}} &
\multicolumn{1}{c|}{\textbf{Local}} &
\multicolumn{1}{c|}{\textbf{Common remarks from}} \\
\cline{2-3}
\textbf{Schemes} & \textbf{DeepSeek-V3} & \textbf{ChatGPT-5} &  \textbf{Simulation}? & \textbf{\hspace{0.6cm} DeepSeek-V3 and ChatGPT-5} \\
\hline \hline
\textbf{CAS-Lock}~\cite{Shakya2020CASLock} & \scorecell{6} & \scorecell{5} & \simcellinc & Self-loop/combinational cycle introduced when rewiring. Cascading of AND-OR not present initially -- corrected with iterative feedback. \\
\hline
\textbf{SARO}~\cite{Alaql2021SARO}     & \scorecell{4} & \scorecell{4} & \Xmark~\textbf{FAIL} & No explicit partition TT transforms. \\
\hline
\textbf{SRLL}~\cite{Hashemi2025SRLL}     & \scorecell{3} & \scorecell{2} & \simcellinc  & Implements random XOR/XNOR locking with key inputs. \\
\hline
\textbf{TriLock}~\cite{Zhang2022TriLock}  & \scorecell{4} & \scorecell{3} & \Xmark~\textbf{FAIL} & EF and thus ESF = ES $\vee$ EF. Mod-$k$/phase gating not implemented. \\
\hline
\textbf{DK-Lock}~\cite{Maynard2023DKLock}  & \scorecell{5} & \scorecell{4} & \simcellfail  & Activation not enforced for $m$ cycles. No global enable gating counter by full key match each cycle. \\
\hline
\textbf{DECOR}~\cite{Hu2024DECOR}    & \scorecell{4} & \scorecell{3} & \simcellinc  & No \texttt{is\_correct}/membership logic to promote keys as correct. \\
\hline
\textbf{Entangle}~\cite{Darjani2022ENTANGLE} & \scorecell{5} & \scorecell{4} & \simcellinc  & Wrong implementation of perturb unit. \\
\hline
\textbf{Harpoon}~\cite{Chakraborty2009HARPOON} & \scorecell{6} & \scorecell{4} & \simcellinc  & No separate obfuscation/authentication modes; just a linear key-sequence gate with reset-on-mismatch. \\
\hline
\textbf{SFLL-HD}~\cite{Yasin2017SFLL}  & \scorecell{3} & \scorecell{5} &  \simcellinc & Not SFLL-HD implementation. Incorrect FSC construction.\\
\hline
\textbf{AutoLock}~\cite{Wang2023AutoLock} & \scorecell{6} & \scorecell{5} & \simcellinc  & Incorrect genotype/candidate definition and fitness function. \\
\hline
\end{tabular}
\end{adjustbox}
\vspace{-.3cm}
\end{table}
%

\subsection{Ablation Studies}

Using the best-performing ChatGPT-5 coder configuration, we conduct an ablation study to evaluate the individual contributions of the different stages in \ours.

\textbf{Content Mining.}
Here, we ablate the content mining process in which the paper-grounded checklist is extracted. We tailor local execution and judge/examiner LLMs to score only by (i) running codes and (ii) cross-checking against the paper without checklists. All other stages remain as is.
As Tab.~\ref{tab:CM_LE_ablation}(left) shows, scheme-specific concepts and structures are missing in the implementation and related checks are failing. Since missions are not explicitly addressed early on,  penalties increase.
\emph{ \underline{Takeaways:} }
Content mining is essential for capturing paper-specific concepts, structures, and behaviour.
Without content mining, reasoning reduces to weak proxies, even for ChatGPT-5. Subsequent evaluation and feedback cannot compensate for such foundational shortcomings.

\textbf{Local Execution.}
Here, we ablate local execution, i.e., the behaviour of the generated code is only judged by static inspection.
Thus, scoring and related agentic feedback acts without the $\mathcal{R}$ criteria.
Once this ablated pipeline of \ours finishes, we run the codes as before, and we use $\mathcal{R}$ again for final scoring.
Results are shown in Tab.~\ref{tab:CM_LE_ablation}(right).
\emph{ \underline{Takeaways:} }
Codes may be structurally sound, i.e., pass all static checks, yet fail in practice.
Local execution exposes related issues and instructs the coder LLM for revisions. Compared to content mining, local execution is less impactful.

\textbf{Independent Examination.}
Here, we ablate the LLM-as-Examiner process.
Thus, final codes are only required to pass against the checklist.
Exemplary results are provided in Tab.~VI.
While the ablated code generation passed (i.e., scored $\geq 8$), some key semantics are missing.
Re-introducing the examiner flagged all these cases and prompted revisions of the checklists.
\emph{ \underline{Takeaways:} } Examination provides paper-anchored semantic checks exposing scoring `blind spots'. These include issues with sequence contiguity, reset semantics, and predicate equality vs. threshold.
 Examination acts as an independent overfitting guard for alignment with the LL schemes. While essential for faithful code generation, its benefits vary across LL schemes with different underlying concepts and complexity.

\begin{table}[tb]
\centering
\scriptsize
\label{tab:exam_ablation}
\caption{Examination ablation induces shortcomings for ChatGPT-5 coding. Initial criteria lists those from $\mathcal{B,C,S,R,P}$ flagged.}
\begin{adjustbox}{width=1.0\columnwidth}
\begin{tabular}{|l|l|l|}
\hline
\textbf{LL Scheme} & \textbf{Initial Criteria} & \textbf{Some T/F questions that had failed} \\ 
\hline
\textbf{SFLL-HD} &
\begin{tabular}[t]{@{}l@{}}
\textbf{B:} equivalence; restore \\cancels flip;  \\[-2pt] wrong-key perturbation \\[-2pt]
\textbf{S:} restore reaches outputs; \\[-2pt]no bypass with wrong key
\end{tabular} &
\begin{tabular}[t]{@{}l@{}}
In the \emph{stripped} design, outputs differ\\[-2pt] from golden iff HD(input,key)$=h$. \\[-2pt]
The HD comparator implements equality \\[-2pt](=$h$), not a threshold. \\[-2pt]
The HD checker is fully wired to designated\\[-2pt] input subsets and corresponding $k$ key bits.
\end{tabular} \\
\hline
\textbf{TriLock} &
\begin{tabular}[t]{@{}l@{}}
\textbf{S:} FSM state count; ES \\[-2pt]drives phase mux/gate-\\[-2pt]enables on the corruption/\\[-2pt]restore path
\end{tabular} &
\begin{tabular}[t]{@{}l@{}}
The required input sequence must appear \\[-2pt]contiguously to trigger ES. \\[-2pt]
On any symbol mismatch, the FSM  \\[-2pt] resets to the start state on the next cycle.  \\[-2pt]
A phase-gating mux enforces mutual  \\[-2pt]exclusivity between ES and the restore/safe path. \\
\end{tabular} \\
\hline
\end{tabular}
\end{adjustbox}
\vspace{-.1cm}
\end{table}

%% file: text/SectionVI-Conclusion.tex
\section{Conclusion}

\ours is a reference-free framework that turns, for the first time, LL papers into executable code.
Addressing this challenge requires more than a simple coding assistant, given the significant semantic gap and domain-specific complexity of LL schemes.
Key stages of \ours are paper-grounded concept extraction and scoring, local execution, and independent final examination.
Ablation studies show that each stage is crucial for guiding LLMs through this complex task.
\ours’s cross-model evaluation improves fidelity of the codes and mitigates potential model-specific bias.
\ours yields faithful artifacts on ten seminal schemes, spanning a diverse range of LL techniques and security philosophies.
All generated codes and results all released at~\cite{lockforge_anonymous_repo}.
Our results show that only recent reasoning models such as ChatGPT-5 reliably meet the thresholds for this complex task, underscoring the timeliness of our work.
Building on this milestone, future work will extend \ours to more schemes and incorporate established LL attacks to further validate resilience of SOTA LL schemes.